# Dynamically twistable three-dimensional moiré photonic crystals

Henri Lemineur,[1] Abhishek Padhy,[1,2] Nicolas Roy, [1,3] Benoit Hackens[2] and Michaël Lobet[1,*]

[1] University of Namur, Department of Physics & Namur Institute of Structured Materials, Rue de Bruxelles 61, 5000 Namur, Belgium

[2]Institute of Condensed Matter and Nanosciences (IMCN), Université Catholique de Louvain (UCLouvain), Louvain-la-Neuve 1348, Belgium

[3] Cenaero, Rue des Frères Wright 29, Gosselies 6041, Belgium

**ABSTRACT**. Three-dimensional woodpile photonic crystals constitute one of the most successful architectures for realizing photonic band gaps, yet their optical response is traditionally fixed by the geometry established during fabrication. Here, we introduce a twist-controlled woodpile photonic crystal in which the relative angular orientation between successive rod layers acts as an additional, *in situ-tunable* geometrical degree of freedom. Using an extension of rigorous coupled-wave analysis adapted to multilayer structures with rotated reciprocal lattices, we systematically investigate the evolution of the transmission spectrum as a function of twist angle. We show that twisting drives the structure through three distinct photonic regimes. In the fully aligned configuration, broad frequency intervals exhibit near-unity transmission. At intermediate twist angles, the spectrum becomes populated by strongly dispersive resonances displaying characteristic Fano line shapes, high quality-factor and pronounced angular sensitivity. As the twist angle approaches 90°, the conventional woodpile structure is recovered, and these resonances evolve into a broad photonic stop band characteristic of three-dimensional photonic crystals. A simplified analytical model based on reciprocal-lattice considerations accurately reproduces the principal resonance modification observed in the numerical calculations. Our results demonstrate a continuous twist-induced transition from broadband transmission to photonic stop bands through an intermediate Fano-resonant regime, establishing twisted woodpiles as a versatile platform for three-dimensional twist-engineered photonics.

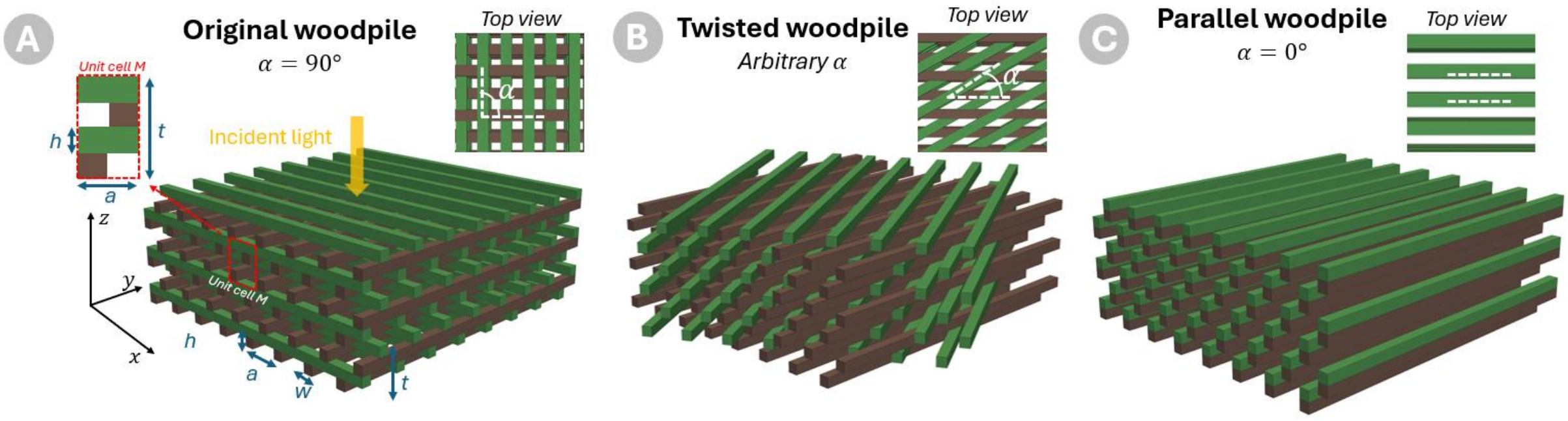

**Fig. 1.** (a) Original woodpile photonic crystal composed of stacked dielectric rods rotated by 90°between successive layers. As in the original woodpile, the 1st and 3rd as well as the 2nd and 4th layers of the unit cell are shifted by half a period. (b) Twisted woodpile geometry investigated in this work, where the relative orientation between adjacent layers is controlled by a continuous twist angle $\alpha$. (c) When $\alpha = 0°$the structure corresponds to fully aligned rods. Note that colors are employed for differentiating odd and even layers, but all rods are made of the same material ($\varepsilon = 13$). Dimensions are $w = 0.3125\ a$, $h = 0.25\ a, t = 4h = a$ with $a$ the in-plane period. $N = 7$ unit cells in $z$ directions are chosen here.

## I. Introduction

The ability to manipulate the flow of light through artificially structured media has been a central objective of photonics since the pioneering works of Yablonovitch[1] and John[2] , who demonstrated that a periodic modulation of the dielectric constant can profoundly alter the propagation of electromagnetic waves and give rise to photonic band gaps. In direct analogy with electronic band theory, photonic crystals enable the tailoring of electromagnetic eigenmodes through geometry alone, leading to a broad range of phenomena including inhibited spontaneous emission, waveguiding, cavity confinement, and enhanced light-matter interactions[3]. Three-dimensional photonic crystals occupy a unique position because they can support complete photonic band gaps, prohibiting propagation in all spatial directions, and for all polarizations [4–6]. Among the various architectures proposed for achieving complete three-dimensional band gaps, the woodpile structure has emerged as one of the most successful and experimentally accessible platforms. Consisting of a multilayer stack of orthogonal dielectric rods (Figure 1a), the woodpile geometry approximates roughly the diamond lattice, a configuration long recognized for its ability to sustain large photonic band gaps[7]. Early microwave experiments demonstrated strong suppression of transmission in periodic rod assemblies, providing the first experimental evidence of three-dimensional photonic band-gap behavior[8]. Subsequent advances in microfabrication enabled the realization of woodpile crystals operating at near-infrared wavelengths[9,10], establishing the structure as a cornerstone of three-dimensional photonics and opening routes toward integrated photonic devices based on fully confined electromagnetic modes.

Despite their remarkable optical properties, conventional woodpile crystals remain fundamentally static. Their photonic response is primarily governed by fixed geometrical parameters such as rod dimensions, lattice spacing, and refractive-index contrast. As a result, their spectral characteristics are typically prescribed during fabrication and are subsequently modified only through defect engineering or changes in the dielectric environment. As photonics shift toward programmable and reconfigurable platforms [11,12], identifying additional geometric degrees of freedom capable of controlling the electromagnetic response of three-dimensional photonic crystals has become an important challenge. Essentially, approaches allowing substantial spectral tuning while preserving the intrinsic advantages of woodpile architecture remain largely unexplored.

A promising route toward geometric control has recently emerged from the rapidly developing field of moiré photonics. Inspired by twisted van der Waals materials[13–15], photonic systems composed of rotated (twisted) or slightly mismatched periodic structures create superperiodic moiré patterns and have revealed a rich variety of phenomena[16], including flat bands[17–19], beam steering[20] and enhanced field confinement[21]. In twisted photonic systems, the relative orientation between adjacent layers provides a powerful degree of freedom for engineering the reciprocal-space landscape. By rotating the constituent lattices, the relative alignment of their reciprocal-lattice vectors is modified, thereby changing the available momentum-matching pathways and the interlayer coupling between photonic modes. This twist-induced modification of the photonic band structure can shift and reshape guided-mode resonances (GMRs)[22,23], whose excitation is governed by phase matching between incident radiation and guided modes. Thus, twist engineering provides a versatile degree of freedom for controlling both the dispersion of photonic modes and their coupling to free-space radiation in periodic media. However, moiré photonics has thus far been explored predominantly in one-dimensional and two-dimensional systems. The possibility of combining the robust band-gap physics of woodpile structures with the geometric tunability of twist-engineered photonic systems remains essentially unexplored. Fundamentally, it is unclear how a continuous rotation between the constituent layers of a three-dimensional photonic crystal reshapes its transmission properties, modifies the formation of photonic stop bands, and influences the appearance of resonant states. Addressing these questions is not only of fundamental interest for wave physics but may also provide new routes toward geometrically programmable photonic materials operating without active tuning elements.

In this work, we introduce a twisted three-dimensional woodpile photonic crystal in which the relative angular orientation between successive layers constitutes a new degree of freedom for controlling electromagnetic transmission (Figure 1). Using an extension of rigorous coupled-wave analysis adapted to multilayer structures with rotated reciprocal lattices [24,25], we systematically investigate the evolution of the transmission spectrum as a function of twist angle. We numerically demonstrate that twisting drives the system through three qualitatively distinct photonic regimes. Near the conventional woodpile configuration, broad photonic stop bands dominate the response, reflecting the characteristic band-gap physics of periodic three-dimensional dielectric crystals. At intermediate twist angles, these gaps fragment into a series of sharp dispersive resonances exhibiting characteristic Fano line shapes and pronounced angular sensitivity. In contrast, approaching the fully twisted configuration gives rise to broad spectral regions of near-unity transmission, revealing a radical transformation of the underlying propagation mechanism. We further show that the resonant states can be captured by a simple analytical model based on reciprocal-lattice considerations and effective-medium arguments. By bridging the historical framework of three-dimensional photonic crystals with the emerging concepts of moiré photonics, our results establish twisted woodpile structures as a versatile platform for twist-engineered light propagation and demonstrate how geometric rotation alone can continuously transform a photonic crystal from a band-gap-dominated regime to a transmission-dominated regime.

## II. Twisted woodpile geometry and numerical framework

The investigated structure is derived from the original woodpile photonic crystal, which consists of a periodic stack of dielectric rods arranged in successive layers with alternating orientations (Fig. 1a). In the standard configuration, each layer is rotated by $\alpha = 90°$ with respect to the previous one and the couple (1st, 3rd) and (2nd, 4th) layers shifted laterally to reproduce the symmetry of a diamond-like lattice known to support pronounced photonic band gaps. This architecture has become one of the most extensively studied three-dimensional photonic crystal geometries owing to its scalability and experimental accessibility.

To introduce an additional geometrical degree of freedom, we replace the fixed $90°$ rotation between adjacent layers by a continuously tunable twist angle $\alpha$ (Fig. 1b). The parameter $\alpha$ therefore controls the relative orientation between consecutive rod layers and continuously interpolates between two limiting configurations. At its minimum value $\alpha = 0°$ (Fig. 1c), all rods become aligned along the same in-plane direction. Intermediate values of $\alpha$ generate a family of twisted woodpile structures characterized by modified reciprocal-lattice symmetries and altered interlayer coupling. The twist angle thus acts as a purely geometrical tuning parameter without changing either the material composition or the overall layer sequence.

The elementary building block is composed of four successive rod layers forming a unit cell $M$. Each rod is characterized by a rectangular cross section of width $w$, height $h$, while the lattice period within each layer is denoted by $a$. Repetition of the unit cell along the stacking direction $z$ generates a finite three-dimensional crystal composed of N unit cells. Throughout this work, all geometrical dimensions are expressed relative to the in-plane period $a$, allowing the results to be presented in a scale-independent form.

The geometrical parameters are chosen to remain compatible with future experimental realization in the microwave X-band while preserving a sufficiently large dielectric contrast to support strong spectral features (here $\varepsilon_{rods} = 13$ and surrounding medium is air). Unless otherwise stated, all simulations are performed using the dimensions summarized in Fig. 1 and numerical details are given in the Supplemental Material.

The optical response of the twisted woodpile is calculated using RCWA. Because successive layers possess distinct in-plane orientations, the conventional RCWA formalism must be extended to account for multiple rotated reciprocal lattices within a single multilayer stack. To this end, we employ a recently developed multidimensional RCWA 4D [20,24,25] approach adapted to structures with incommensurate or rotated periodicities.

For each value of the twist angle $\alpha$, transmission spectra are computed under normal incidence over the frequency range of interest (0.30-0.55). The resulting calculations provide a continuous description of the evolution of transmission spectrum as a function of twist angle and form the basis of the analysis presented in the following sections.

## III. Results

### From Transparency to Photonic Stop Bands

Figure 2 displays the calculated transmission spectrum of the twisted woodpile structure as a function of reduced frequency $\frac{\omega a}{2\pi c}$ and twist angle $\alpha$. The map reveals a remarkably rich evolution of the optical response as the relative orientation between successive layers is continuously varied from the fully aligned configuration ($\alpha = 0°$) to the conventional woodpile geometry ($\alpha = 90°$). Rather than producing a simple shift of spectral features, the twist fundamentally reorganizes the spectrum, giving rise to three distinct photonic regimes. Near $\alpha = 90°$, the transmission is dominated by broad spectral regions of strong attenuation characteristic of a photonic stop band. These features originate from Bragg scattering within the three-dimensional woodpile lattice and reproduce the well-known optical response of original woodpile photonic crystals[7].

Here, the normalized bandgap is $\frac{\Delta\omega}{\omega_0} = 37.3\%$. As the twist angle is reduced, the stop band progressively fragments, and a dense set of narrow dispersive resonances emerges. Their frequencies evolve continuously with $\alpha$, forming the intricate spectral branches visible in Fig. 2. Continuous vanishing of the bandgap size is presented in supplementary materials. Upon further reduction of the twist angle, the resonant structure gradually disappears, and the spectrum becomes dominated by broad transmission windows. Near $\alpha = 0°$, transmission approaches unity over extended frequency intervals, indicating a profound modification of the underlying scattering mechanisms.

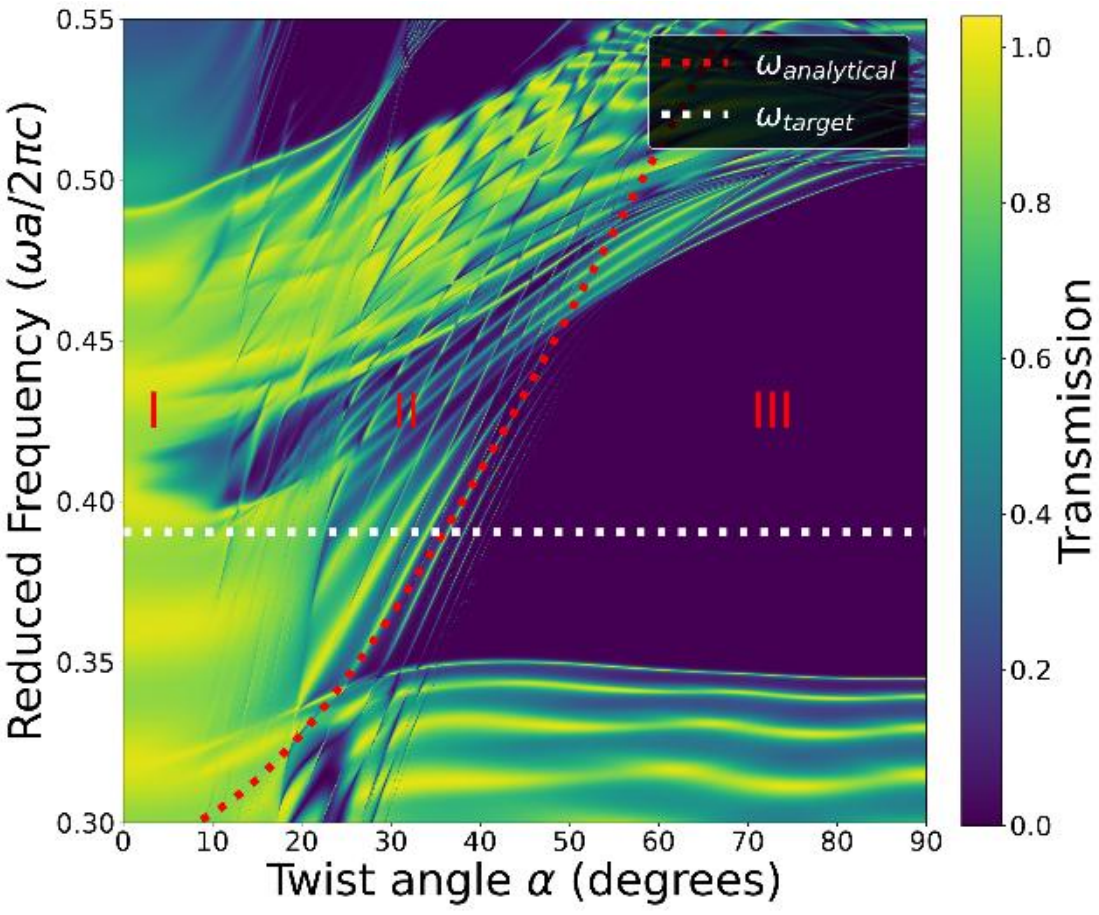

**Fig. 2** Transmission map as a function of the twist angle $\alpha$. Red line corresponds to the analytical model of eq. (1) and white line to the target frequency shown in figure 3.

The transmission map therefore reveals a continuous transition between three regimes, labelled I-III in Fig. 2. Regime I corresponds to the fully aligned structure and is characterized by broadband transmission. Regime II is dominated by strongly angle-dependent resonances, many of which exhibit the characteristic signatures of Fano interference. Regime III corresponds to the conventional woodpile limit, where broad stop bands govern the optical response. The boundaries between these regimes are gradual rather than abrupt, highlighting the continuous nature of the twist-induced spectral evolution.

The red dashed curves superimposed in Fig. 2 denote the resonance frequency predicted by the analytical model introduced below. Despite its simplicity, the model reproduces the dominant dispersive trend observed in numerical calculations and correctly captures their dependence on twist angle. The overall agreement suggests that the evolution of the resonant spectrum is primarily governed by the geometrical modification of the reciprocal lattice induced by the twist.

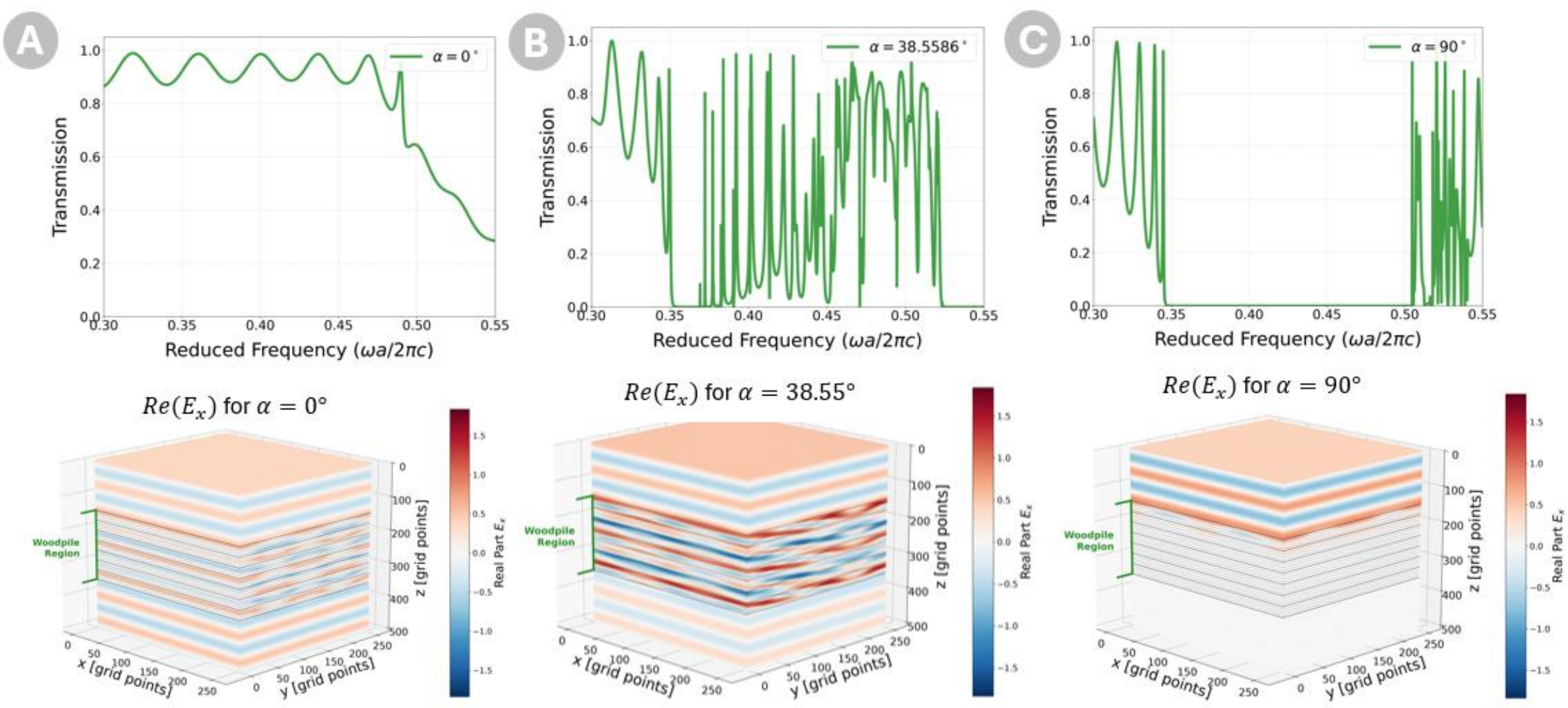

**Fig. 3** Transmission spectra and corresponding field maps for (a) $\alpha = 0°$ - Regime I full transmission, (b) $\alpha = 38.55°$ - Regime II moiré physics and (c) $\alpha = 90°$ - Regime III photonic stop band. Note that light is incident from the top of the woodpile and the frequency of operation is $\omega = 0.392$.

### Physical origin of the three regimes

To elucidate the physical mechanisms responsible for the twist-dependent spectra presented in Fig. 2, Fig. 3 displays representative transmission spectra together with the corresponding electric-field distributions for the three characteristic regimes identified above, here chosen at $\omega_{target} = 0.392$.

The first regime occurs for small twist angles, where the transmission spectrum exhibits broad frequency intervals with transmission approaching unity with Fabry-Perot oscillations type [Fig. 3(a)]. In this limit, successive rod layers become nearly aligned and the structure departs significantly from the conventional woodpile architecture. Consequently, the three-dimensional Bragg-scattering processes responsible for stop-band formation are strongly suppressed.

The corresponding field distribution confirms this interpretation. Instead of being localized within specific regions of the crystal, the electric field extends throughout the entire structure and maintains a significant amplitude across all layers. The incident wave therefore propagates efficiently through the finite crystal, explaining the broad transmission windows observed in the spectrum.

A qualitatively different behavior emerges at intermediate twist angles. As seen in both Fig. 2 and Fig. 3(b), the broad transmission windows progressively fragment into multiple narrow resonances whose frequencies evolve continuously with angle. Many of these resonances exhibit pronounced asymmetric line shapes characteristic of Fano interference.

The corresponding field maps reveal strong localization inside the structure, with energy concentrated within specific layers and regions of the twisted lattice. Such localization is absent in the highly transmitting regime and indicates the excitation of quasi-guided resonant states weakly coupled to free-space radiation. The observed Fano profiles therefore originate from the interference between these localized states and the continuum of propagating modes supported by the structure, reminiscent of what occurs to their 2D counterpart[26]. The strong angular sensitivity of the guided resonances demonstrates that twisting provides a particularly efficient mechanism for controlling their spectral position[22,23,27].

As the twist angle approaches 90°, the resonant branches gradually merge into broad regions of suppressed transmission. The spectrum progressively recovers the characteristic response of conventional woodpile photonic crystals, with pronounced stop bands spanning an extended frequency range [Fig. 3(c)]. The associated field distributions reveal strong attenuation within the structure, with only weak field penetration beyond the first layers. This behavior is consistent with the multiple Bragg-scattering processes responsible for the formation of photonic stop bands in periodic three-dimensional dielectric lattices [3,7,8]. The structure therefore transitions smoothly from the aligned transmitting configuration to the well-established woodpile limit as the twist angle increases.

The physical origin of the resonances can be understood using a simplified analytical model based on the reciprocal-space representation of the twisted lattice[22]. Rotating successive layers modifies the set of reciprocal lattice vectors contributing to diffraction and introduces additional coupling channels that are absent in the original woodpile geometry. The resonance frequencies can then be estimated by combining these geometrical considerations with an effective dielectric permittivity describing the multilayer structure giving the approximation:

$$\omega_{approx}(\alpha) = \frac{\frac{2\pi c}{a}\sqrt{5-4cos\alpha}}{\sqrt{\varepsilon_{eff}}} \quad (1)$$

with $\varepsilon_{eff}$ being the effective permittivity (supplementary materials for more details). The resulting prediction is shown as dashed curves in Fig. 2. Despite the simplicity of the model and the absence of adjustable geometrical parameters beyond the effective permittivity, the calculated resonance positions closely follow the dominant trend obtained from the full RCWA 4D simulations. The agreement indicates that the principal spectral features are governed primarily by the geometrical evolution of the reciprocal lattice induced by the twist.

Taken together, the field distributions and analytical model provide a unified interpretation of the spectral evolution observed in Fig. 2. The aligned configuration supports extended transmitting states, the conventional woodpile exhibits Bragg-induced stop bands, and intermediate twist angles enable the formation of localized resonant states that mediate the continuous transition between these two limiting regimes.

## IV. Conclusion

In this work, we investigated the evolution of the optical response of a three-dimensional woodpile photonic crystal as a function of the relative twist angle between successive layers. Using an RCWA framework adapted to rotated multilayer geometries, we showed that twisting provides a powerful geometrical degree of freedom capable of continuously reshaping the transmission spectrum without modifying either the material composition or the lattice dimensions.

The transmission map of Fig. 2 revealed a continuous evolution between three distinct regimes. In the fully aligned configuration, the structure exhibits broad transmission windows with transmission approaching unity over extended frequency intervals. At intermediate twist angles, the spectrum becomes dominated by narrow dispersive resonances displaying characteristic Fano line shapes with high Q-factor and strong angular sensitivity. As the twist angle approaches the conventional woodpile limit, these resonances gradually evolve into broad photonic stop bands characteristic of three-dimensional woodpile photonic crystals.

Field distributions and analytical modeling provide a unified interpretation of this behavior. The high-transmission regime is associated with extended propagating states, whereas the intermediate regime originates from quasi-guided resonances supported by the twisted lattice analogue to the 2D case. A simplified reciprocal-space model enables to recover the main trend of the twist-angle dependency, demonstrating that the spectral evolution is primarily governed by the geometrical modification of the lattice induced by the twist.

Beyond the specific woodpile architecture considered here, our results establish twist engineering as a versatile strategy for controlling the optical properties of three-dimensional photonic crystals. The ability to continuously transform a structure from a stop-band-dominated regime, through a resonant regime, and ultimately into a highly transmitting configuration using a single geometrical parameter opens promising perspectives for tunable filtering, wavelength-selective devices, and reconfigurable photonic architectures. Experimental confirmation in the microwave regime is a privileged road because of cm scale dimensions. More broadly, these findings extend concepts originally developed in moiré photonics to fully three-dimensional systems and provide a foundation for future experimental investigations of twisted photonic crystals.

## Supplementary materials

## Evolution of the band gap with twist angle

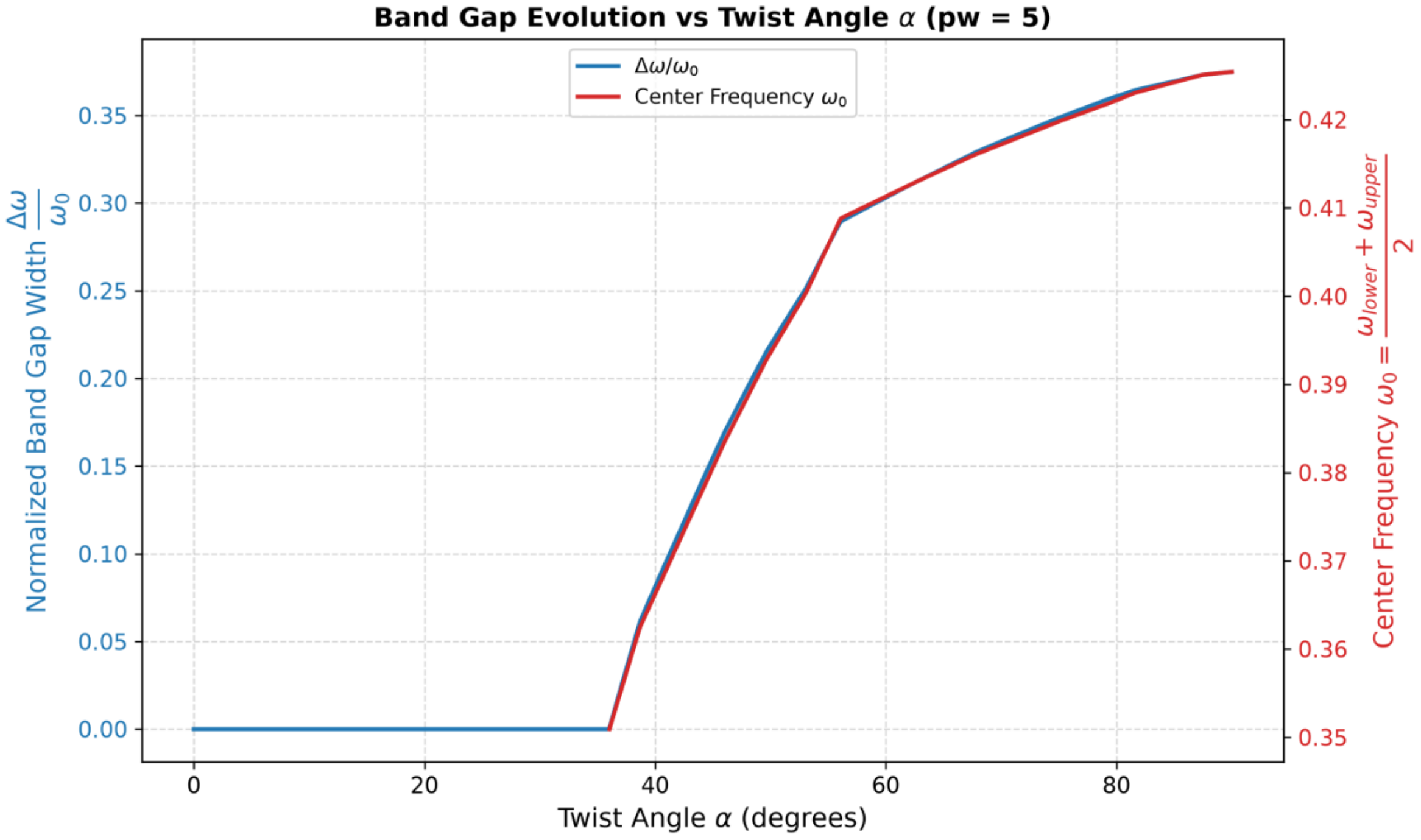

**Figure S1**: numerical evaluation of the normalized band gap width $\Delta\omega/\omega_0$ with $\omega_0$ being the central frequency as a function of the twist angle (blue curve). Note that the central frequency $\omega_0$ is calculated as the average frequency between the upper and lower frequencies of the bandgap ( $T = 0$ ) which is also dynamically moving with the twist angle.

## Derivation of the analytical approximation for the resonance frequency

To estimate the characteristic resonance frequency $\omega_{approx}$, we use a simplified analytical model based on the effective dispersion relation of the guided modes[22]. Under normal incidence, the incident wave couples to the guided modes via the reciprocal lattice vector of the twisted structure. For the specific resonance described in Equation (1), the relevant reciprocal vector is $g = 2b_1 - b_2$, which corresponds to the diffraction orders $n = 2$ and $n' = -1$ . The norm of this Moiré vector evaluates to $|g| = \frac{2\pi}{a}\sqrt{5 - 4cos\alpha}$.

The resonance frequency is approximated using the dispersion relation $\omega = \frac{c|k|}{\sqrt{\varepsilon_{eff}}}$, where $|k| = |g|$ at normal incidence. Other branches could be estimated by a similar treatment.

To calculate the effective permittivity $\varepsilon_{eff}$ of the structure, we approximate a single layer of the woodpile as a 1D photonic layer. The effective permittivity is then computed as a volume average: $\varepsilon_{eff} = f\,\varepsilon_{rods} + (1 - f)\,\varepsilon_{background}$, where $f = 0.3125$ is the filling factor of the rods within the layer, and $\varepsilon_{rods} = 13$ and $\varepsilon_{background} = 1$ are the permittivities of the rods and of the surrounding air, respectively, giving $\varepsilon_{eff} = 4.75$. This arithmetic averaging is strictly valid for the polarization with the electric field parallel to the rods; for the orthogonal polarization, a harmonic average ($\frac{1}{\varepsilon_{eff}} = \frac{f}{\varepsilon_{rods}} + \frac{1-f}{\varepsilon_{background}}$) would be more appropriate, and the two estimates differ substantially given the strong anisotropy of the structure. This choice of a single scalar effective permittivity is therefore an approximation, retained here for the purpose of a first-order analytical estimate. Substituting the norm of the reciprocal vector g and this effective permittivity into the dispersion relation yields the approximation presented in Equation (1).

**Numerical convergence test and computational efficiency**

In the Rigorous Coupled-Wave Analysis (RCWA) method, the accuracy of the electromagnetic field calculations depends directly on $N$, the number of plane waves used to truncate the Fourier expansion. To ensure physical accuracy while preventing prohibitive computational costs, a convergence study was conducted.

The transmission and reflection spectra of the structure were simulated at a fixed twist angle for increasing values of $N$. We evaluated the numerical convergence by quantitatively tracking three critical spectral indicators:

1. The frequency position of the photonic band gap.
2. The spectral width of the band gap.
3. The position of the first resonance peaks immediately preceding and following the band gap.

As the number of plane waves increases, the frequencies of these features asymptotically approach a stable limit. Convergence is considered achieved when the variation of these reference positions becomes negligible between successive increments of $N$.

However, because the computational time scales rapidly with the number of plane waves, the stabilization of the spectral features was carefully weighed against the simulation duration. Consequently, $N = 5$ was selected, as it was deemed the optimal compromise between sufficient physical accuracy and practical execution time for the simulations.

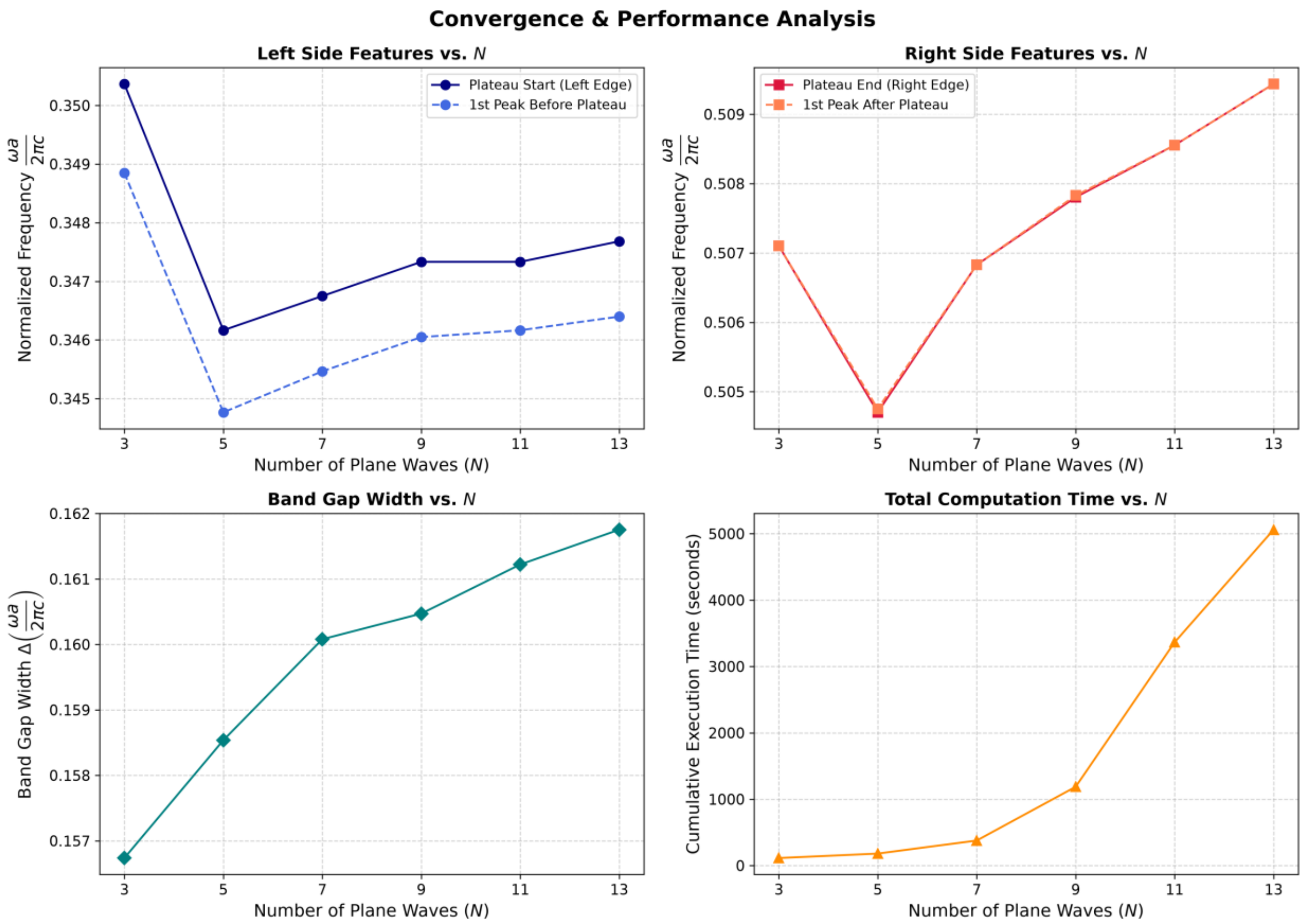

**Figure S2**: convergence study: evolution of the different figure of merit.